\documentclass[a4paper,12pt]{article}
\usepackage[utf8]{inputenc}
\usepackage{graphicx} 
\usepackage{caption}
\usepackage{subcaption}
\usepackage{multicol}
\usepackage{amsmath}
\usepackage{amsfonts}

\normalbaselines
\title{Towards a Framework for Social Mechanics}
\author{VS Morales-Salgado}
\date{}

\begin{document}

\maketitle

\begin{abstract}
Social physics explores the possibility that mathematical structures developed in physics may provide useful descriptions of certain social phenomena. 
In this work, we propose an effective mechanical framework for modelling social change in terms of positions in a space of social stances, together with concepts analogous to motion,
inertia, interaction, and force. 
A central feature of the framework is the introduction of position-dependent inertial responses, allowing susceptibility to social change to vary across stance-space. 
Within this setting, we investigate deterministic and stochastic models of social evolution, including free motion, effective interactions, and diffusion-driven dynamics. 
We also discuss Lagrangian and Hamiltonian formulations associated with the proposed framework. 
As an illustrative application, we model partisan preference distributions in United States presidential elections through effective drift and diffusion processes. 
The framework is intended as a phenomenological and exploratory approach to social dynamics rather than as a fundamental description of human behaviour.
\end{abstract}

\section{Introduction}\label{sec1}
The objective of this work is to employ the key idea behind social physics that, under appropriate conditions, the tools developed to describe physical systems can be used to describe certain social systems \cite{w69,j22}.
We focus on fundamental mechanical concepts to contribute towards understanding social change, its nature and its causes.
Even more, establishing appropriate analogues of mechanics in social phenomena  contributes to further explore connections of other physical concepts in a systematic way.
This may be so since mechanical concepts are foundational to other areas of physics like thermodynamics, quantum mechanics, and general relativity, given the appropriate mechanisms like statistical mechanics, quantization and covariant formulations, respectively.
 
We are looking for a description of social phenomena in terms of stances about a social question (position) \cite{hp87,g05,s11}, social change (motion) \cite{e73,o14,s19}, predisposition to change (inertia) \cite{p69,b85,zsmb12,b13}, advocacy and compellingness (forces and interactions) \cite{j87,k11}, and others in a similarly fashion.
However, we expect the analogy not to be perfect, but a useful approximation \cite{sb37,l19}.
We also expect there to be differences in the relations among them compared to what happens in physics.
In other words, one would expect that a possible mechanical theory of social change be a genuine \emph{social mechanics} in its own right.
 
In a substantial part, this proposal aims to provide a mathematical framework to observations about social change like the following, taken from \cite{s19}:
    ``It is important to emphasize that with small variations in starting points
    and inertia, resistance, or participation at the crucial points, social change
    may or may not happen. Suppose that a community has long had a norm
    in favor of discrimination based on sexual orientation; that many people in
    the community abhor that norm; that many others dislike it and that many
    others do not care about it; that many others are mildly inclined to favor
    it; and that many others firmly believe in it. If norm entrepreneurs make a
    public demonstration of opposition to the norm, and if the demonstration
    reaches those with relatively low thresholds for opposing it, opposition will
    immediately grow. If the growing opposition reaches those with relatively
    higher thresholds, the norm might rapidly collapse. But if the early public
    opposition is barely visible or if it reaches only those with relatively high
    thresholds, it will fizzle out and the norm might not even budge''.
This considerations show similarities between evolution of social systems and that of physical systems.
Indeed, the coincidences mostly distil from the more fundamental concept of change with respect of time.
 
Indeed, the work presented here belongs to the realm of quantitative methods and mathematical models for human and social research \cite{s78}.
As such, it does not give a fundamental explanation of why such systems behave the way they do.
What it does is to provide tools to model this behaviour to help investigate its nature through measurable traits.
We do not pretend to substitute already existing theories about social behaviour, but to aid them by introducing quantitative tools based on well known results about analogous models in physics.
 
We are confident that the results will prove interesting to both social researchers, as well as physicists.
On the one hand, investigators of social phenomena can put these results to test while also using them to characterize individuals and populations, their traits and conditions,  as well as the possible relations among all of them.
On the other hand, we believe that physicists will find that the resulting generalizations of physical concepts yield novel and fascinating possibilities to further develop them.

The rest of the article is organized as follows: 
In Section \ref{FC} we introduce and develop fundamental concepts for a mechanical theory of social change analogously to classical mechanics in its Newtonian formulation.
In Section \ref{MT} we use those fundaments to describe some initial models. 
Specifically, we study the motion of a particle subject to free motion, a force described by Hooke's law, and a stochastic interaction, governed by the Smoluchowski equation.
As an example of the use of this last one for modelling real data, we describe partisan preferences in the United States as outlined by presidential elections.
In Section \ref{Other} we briefly discuss how other formulations of mechanics, known to be equivalent to the famous Newton's laws can be understood in the light of our proposal.
Finally, in Section \ref{CS} we present the concluding remarks of this investigation, as well as a some prospective lines to continue this research.

\section{Fundamental concepts}\label{FC}
In this section we will present the concepts that will allow us to build a mechanical theory of social change.
A particle in our social models represents an single conviction held by a unit of the social universe.
Just like in physics, here a particle is an idealization \cite{g14}.
It helps to describe the simplest component of matter, here understood as a society, and upon aggregation with other particles yields more complicated forms of social matter.
Actually, this illustrates a powerful trait of physics: it departs from simple cases to be aggregated non-trivially to explain more complex phenomena.
 
Since societies are composed of individuals, a particle can be readily identified with the convictions of an individual person.
However, other realizations of a particle could be made \cite{y93}. 
For example, a social group fairly homogeneous with respect to a trait of interest could be identified with a particle, just like the centre of mass of an extended object, e.g. a ball, can be identified with a particle, although it is composed of more fundamental particles.

\subsection{Position}
The first concept that we wish to discuss is that of \emph{position} or, more generally, \emph{configuration}. 
The position of a social system should describe its status with respect to an attribute of which change (or lack thereof) occurs.
More concretely, a positions can be identified with beliefs, convictions or stances about a social question.
Then, a position in social mechanics is treated here in the sense of the common phrasing ``what is your position on this topic?---My position on it is...'' and others similar to these.
 
We also require the position to be describable by coordinates in a space.
Thus, this \emph{space} is generally the set of all possible positions an individual can take on a topic.
More specifically, the system at hand shall give us hints on the mathematical properties of such a space.
For brevity, we can think of this space as a subset of $\mathbb{R}^n$, that is, we can specify a position by $n$ real coordinates in the usual notation $(x_1,x_2,\dots,x_n)$. 
 
This picture has some underlying assumptions about the type of characteristics that can be modelled by it.
We suppose that in the space of social positions we can assign univocally a set of coordinates $(x_1,x_2,\dots,x_n)$ to a given stance.
That is, we can somehow measure stances or, at least, there is a proxy way of measuring them.
We also suppose that there is a ``well-behaved'' way of comparing different stances.
This is a matter that is readily found in the fundaments of some already existing conceptual frameworks around social phenomena, like in the utility theory of economics.
    
\subsection{Motion}
Since we wish to describe \emph{motion}, i.e. evolution or change with respect to time $t$, each $x_i$, where $i=1,\dots,n$, is in general a function of time: $x_i=x_i(t)$. 
For example, consider the somewhat common affirmation that ``fashion is cyclical''.
This statement refers to a social trait: a stance on a way of dressing, but it also depicts heuristically an observation about how this trait evolves in time.
Even more, it claims a change in time very much familiar to physics: an oscillatory motion.
Such a model of a trait varying in time is not new in social sciences and should be understood closely related to the method of analysing time series. 

\begin{figure}[h]
     \caption{Example of a possible evolution of a stance $x=x(t)$.}
     \centerline{\includegraphics[width=\linewidth]{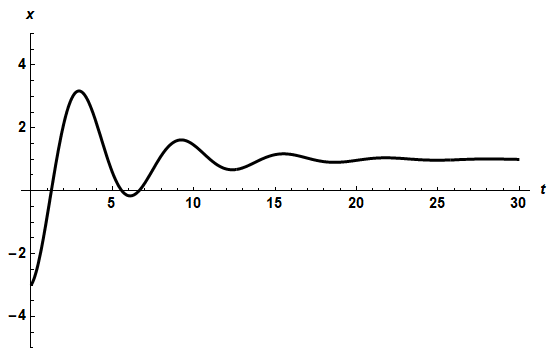}}
     \label{Ex1}
    \end{figure}

To illustrate motion in social mechanics, let us suppose that $n=1$, then we only need one number or coordinate $x$ to describe the stance of an individual or group of individuals regarding an issue like, for instance, a policy.
As an example, this evolution could look like in Figure \ref{Ex1}, where a positive $x$ means an stance favourable towards the policy, the more positive, the more favourable; a negative $x$ implies an unfavourable view of the policy, the more negative, the more critical; and $x=0$ is a position of indifference towards the policy. 

We can see that the exemplified evolution describes a strong initial negative stance on the issue, followed by a change towards a strong positive view and further oscillations the settle in a rather positive althoug moderate stance.
The heuristics behind the example is an agent exploring different stances, possibly in response to new evidence, maturing towards a definite position on the topic.  
Of course, we have assumed a continuous behaviour of the stance with respect to time.
However, this is an illustration and assumptions can be dropped or added according to the phenomena of interest.
 
Note that a trivial state of motion is possible: no motion at all, also called \emph{rest}.
In fact, we can see that the particle described in Figure \ref{Ex1} tends towards rest as time goes by.
 
\subsection{Inertia}
So far the concepts we presented have not been particularly heavy on physics. 
Position as a stance on social issues and motion as a a change in position over time have been used in social sciences by other names in a transparent way.
Now, we proceed to a concept that is in the kernel of mechanical treatments in physics: inertia.
 
Inertia is the tendency to maintain a specific state of motion, usually called \emph{free motion}.
Conversely, this tendency can be understood as an opposition to change the state of motion in response to efforts to modify it.
Upon observations of different systems we can assign a parameter quantifying said opposition.
We call it \emph{inertial mass} or just mass, for now.
This mass, usually denoted by $m$, provides us with a powerful quantifiable property to characterize how individuals, or groups of them, respond to change.
In physical heuristics, there are individuals that are heavier or lighter than others depending on how difficult or easy it is to provoke a change in their motion, respectively.
 
From Newton's first law of mechanics, the state of free motion in physics is the uniform linear motion.
In turn, this is closely related to the idea of particles having a constant inertial parameter, that is, mass.
However, there is no reason for this to be the case for social phenomena.
In fact, we shall see that it makes sense to have \emph{social masses} that depend on the position of the particle.
This is one of the interesting generalizations resulting from applying physical ideas to social studies.

It is important to distinguish between the general meaning of mass in the present  framework and the particular dynamical behaviours resulting from specific mass functions. 
In its most general sense, mass characterizes the resistance of a social particle to changes in its state of motion. 
Since the mass may depend on position, this resistance can itself vary across the space of stances, allowing different susceptibilities to change depending on the current configuration of the system.

Specific choices of position-dependent mass functions may additionally produce effective restoring, confining, or repulsive behaviours around particular regions of the configuration space. 
Such behaviours, however, are not fundamental definitions of mass itself, but emergent dynamical consequences of the particular model under consideration.

There is another quantity closely related to mass. 
It is the \emph{momentum}, usually denoted by $p=m\,\frac{{\rm d}x_i}{{\rm d}t}$, that is, the product of the mass $m$ and the  velocity $\frac{{\rm d}x_i}{{\rm d}t}$, which is the rate of change of the position $x_i$ in time, obtained by deriving $c_i$ with respect to time $t$.
It is closely related because, as we shall see, it is the quantity upon which externalities act more directly and transparently.
The reason being that $p$ encompasses, by definition, both the state of motion and the opposition to change it.
Thus, when we refer to the state of motion we mean a given momentum $p$.
 
\subsection{Interaction}
Mass is a property associated with the social unit susceptible of motion, even if it depends on its position.
On the other hand, we still need to describe externalities, that is, the action of others on the particle.
We do this through the concept of \emph{force}.
Forces are in general functions of position and time: $F=F(x_i,t)$, where $i=1,\dots,n$ and the specific mathematical expression is proposed and confirmed for each social phenomena.
 
In line with previously introduced concepts, a force is an influence causing a change in the state of motion of a particle. 
Similarly, here an \emph{interaction} is the exercise of forces among several particles.
We say ``here'' because in physics there are other formulations of mechanics.
For now, we shall stick to the use of forces since they are valuable in explaining intuitively the comparison between physical and social phenomena.
 
A concept closely related to motion and force is that of \emph{equilibrium} in that it is observed as a lack of change in the motion due to the exact balance of opposing forces.
Indeed, the lack of a net resulting force on a particle is not rest but the continuance of its current motion.

It is important to emphasize that, in the present framework, a social force is not assumed to be a fundamental entity independently measurable from social dynamics.
Rather, it should be understood as an effective quantity inferred from systematic changes in observed social trajectories. 
In this sense, forces play a role analogous to effective interactions in phenomenological physics: 
they provide a mathematical description of how social influences manifest through changes in the evolution of stances, without necessarily specifying the underlying mechanisms producing them.

Before we move on, a caveat is at hand. we have described position as points in a space of possible stances and, with it, we have disregarded physical space, that is geographical separation between individuals and societies.
So, how are they interacting? 
Certainly, there must be a physical channel carrying interactions. 
Whether it be in-person conversations, interactions through social media, action through an intermediate individual, or other ways, it does not matter for our current formulation.

In this work we are concerned about the effect of the interaction on the stances of each participant in a social group and with a given social environment.
We do not mean to say that physical distance is unimportant. 
Rather that we assume that a channel for communicating interactions is already established of which the physical separation, as well as other properties, form the set of parameters shaping the interaction in the space of social positions.

The framework proposed here does not assume that social systems are fundamentally mechanical in nature; rather, it explores whether mechanical structures can provide useful effective representations of certain aspects of social evolution.
 
Modernity is key to the relevance of this framework. 
In the past, when social systems contained a smaller population, the exchange of ideas only occurred when people physically encountered each other and in a slower rate.
Nowadays, there is a bigger, faster and more interacting community.
There is also more data available to be understood on itself, but also to better understand social change, in general.

\section{A mechanical theory of social change}\label{MT}
Here we put all the ideas from the past section to work by building mathematical relations among them. 
We also derive some consequences and develop intuition through some general examples.
 
\subsection{The motion of a particle}\label{parmot}
Again, analogously to classical mechanics, we rely on an idealization of the minimal unit of study.
In physics it is the \emph{particle}, while for social systems this could be an individual person, for example. 
As in physical mechanics, the notion of free motion employed here should be understood as an idealized reference dynamics rather than as the description of a completely isolated real system. 
Human agents are continuously subject to multiple social influences and environmental factors. 
However, introducing a notion of free motion provides a useful baseline against which effective interactions and deviations from unperturbed evolution can be characterized mathematically.

Thus, from the conceptual considerations discussed in previous sections, we may model the evolution of an individual through the effective equation in terms of momentum $p$:
    \begin{equation}
     F = \frac{{\rm d}p}{{\rm d}t} \,.
    \end{equation}
Thus, within this framework, forces are operationally identified through changes in the state of motion of social trajectories.

This realizes the proposed framework since it relates the net effective force $F$ to changes in the state of motion $\frac{{\rm d}p}{{\rm d}t}$. 
It is rather convenient too, because the momentum helps describe the tendency of a change in the state of the individual as well as a way to compare the opposition to change between different individuals. 
This is clearer in the known presentation of mechanical momentum as $p=mv$, where the velocity is given by $v=\frac{{\rm d}x}{{\rm d}t}$.
Thus, we can write
    \begin{equation}
     F = \frac{{\rm d}}{{\rm d}t}\left(m\,\frac{{\rm d}x}{{\rm d}t}\right) \,.
    \end{equation}
    
Now, some common simplifying assumptions from physics are not so immediately applied in human behaviour scenarios. 
For example, while inertial mass in ordinary mechanics is typically assumed to be positive, effective social inertial parameters need not satisfy the same restriction a priori.
In particular, some forms of overreactive or oppositional responses to external influences may be represented phenomenologically through effective negative values within specific models.
As we will further see, many other assumptions common to physics can be also dropped, leading to models that hopefully result interesting to physicists as new playgrounds for their methods; as well as for social and behavioural scientists, as new tools emerge.
    
As another example of dropped assumptions, consider that, for human behaviour it is difficult to imagine that resistance to change, represented by inertial mass, does not depend on the particle's configuration.
More specifically, we consider the general case where mass is a function of position: $m=m(x)$.
Thus, the effective evolution equation looks like follows:
     \begin{equation}\label{2ndlaw}
      F = m\,\frac{{\rm d}^2}{{\rm d}t^2}x 
       + \left(\frac{{\rm d}}{{\rm d}x}m\right)\left(\frac{{\rm d}}{{\rm d}t}x\right)^2 \,,
     \end{equation}
where we have used the notation ${\rm d}_t=\frac{{\rm d}}{{\rm d}t}$ and ${\rm d}_x=\frac{{\rm d}}{{\rm d}x}$.
We shall use this notation from now on, as well as its partial derivative forms $\partial_t$ and $\partial_x$.
Therefore, a key proposal of this work, in terms of a mathematical formalism, is that individual evolution might be modeled as particles with position-dependent masses.

Equation (\ref{2ndlaw}) should be understood as an effective low-order dynamical model rather than as a fundamental law of social behaviour. 
The second-order structure is adopted here as the simplest nontrivial extension of mechanical motion capable of encoding both inertial effects and interactions. 
More general descriptions may involve higher-order derivatives, memory effects, nonlocal interactions, or stochastic contributions, depending on the social phenomena under consideration.
    
Since the effective force $F = F(x,t)$ parameterizes external influences acting on the individual, the first term in the right-hand-side explains the usual: 
a greater force $F$ produces a greater change in the state of motion of the particle $\frac{{\rm d}^2}{{\rm d}t^2}x$. 
But, also, a greater inertia $m$ requires the application of a greater force $F$ to obtain a given change in motion.

The second term in the right-hand-side also yields a relation between the magnitude of the force $F$ and a change in the state of the individual $\frac{{\rm d}}{{\rm d}t}x$. 
However, this time the change is in position and the relation is quadratic, and the proportionality between them is given by the rate of change of the mass with respect to the position.
Thus, it is increasingly harder for a forcing agent to raise the speed of change, irrespective of the direction. 
Nonetheless, this is mediated by the the way the inertia of the individual depends on the stance of the individual.
Arguably, such a model is reasonable for changes of individuals interacting in a society.
    
A first interesting case of equation (\ref{2ndlaw}) is when there is no force: $F=0$, and whose general solution is implicitly given by
        \begin{equation}
         x + c_1 = \int_1^x \frac{m(s)}{c_1} \,{\rm d}s \,.
        \end{equation}
This represents an idealized evolution in the absence of explicitly modeled
external interactions.
Of course, just like in physics, this is an idealized scenario, although very useful as an initial system.
Particles, whether physical or social, are very often subject to forces changing their motion.
    
For illustrative purposes, consider a mass function $m(x)=x^2$.
As a positive-concavity function, it is a reasonable position-dependent mass for the type of phenomena we are trying to study, since such a mass describes individuals that move freely around a prescribed position consistent with a locally stable stance, here denoted by $x=0$.
Nonetheless, the ultimate test for a model is its comparison with measurements. 
Any other mass function adequate with systematic observations can be used.
    
There is another reason (well known to physicists) to use $m(x)=x^2$ as a starting example: approximations by power series of more complicated mass functions can be truncated to second order and still reveal a vast information about the system.
    
In any case, let us proceed with the example by considering the equation of free motion:
        \begin{equation}
         0 = x^2\,{\rm d}_t^2 x + 2x\left({\rm d}_t x\right)^2 \,.
        \end{equation}
The solution to this equation is
        \begin{equation}
         x(t) = c_2\,(c_1 + 3t)^{1/3} \,,
        \end{equation}
where $c_1$ and $c_2$ are constants fixed by the initial conditions of the system.

\begin{figure}[h]
     \caption{Free evolution of a system with a squared position-dependent mass and different initial conditions.}
     \centerline{\includegraphics[width=\linewidth]{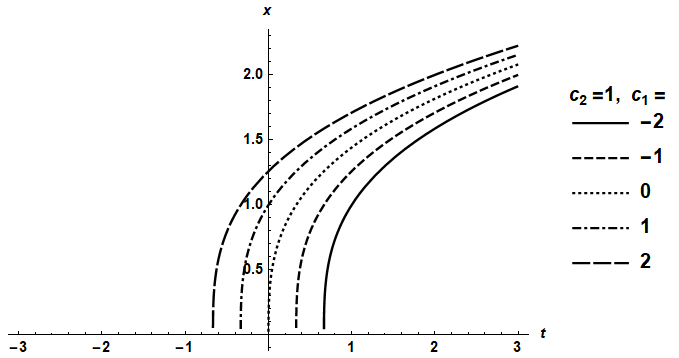}}
     \label{free1}
    \end{figure}

\begin{figure}[h]
     \caption{Free evolution of a system with a squared position-dependent mass and different initial conditions.}
     \centerline{\includegraphics[width=\linewidth]{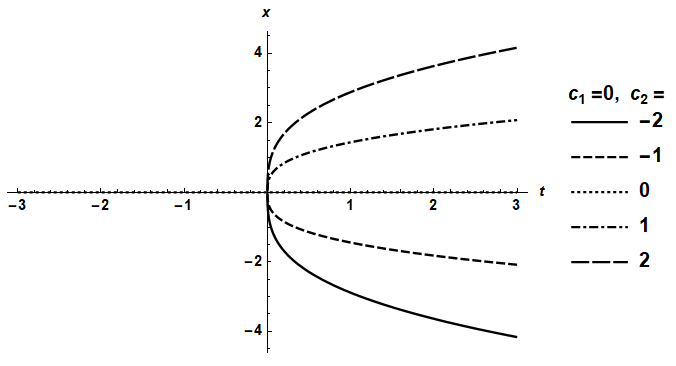}}
     \label{free2}
    \end{figure}

Figure \ref{free1} shows examples of curves describing the  evolution in time resulting from varying the first initial condition, $c_1$, while keeping the second one, $c_2$, fixed.  
Figure \ref{free2} shows examples of curves describing the  evolution in time resulting from fixing the first initial condition, $c_1$, while varying the second one, $c_2$.  
Notice that in the Figure \ref{free2} one can find the simplest scenario: no motion at all. 
This can be seen here for the case where $c_1=c_2=0$, that corresponds to a static undisturbed system.
    
Trajectories are similar for other concave up parabolic mass functions of the form $m=a(x-x_0)^2+m_0$. 
These result from translating the vertex from the origin $(0,0)$ to the point $(x_0,m_0)$.
Thus, they model individuals with different minimal value of mass $m_0$ at a characteristic position $x_0$.
Parameter $x_0$ can be understood as the stance where an individual moves with the most freedom, while parameter $m_0$ can be seen as a measure of how difficult it is to move the individual from that position. 

\begin{figure}[h]
     \caption{Concave up parabolic mass functions of three particles with different.}
     \centerline{\includegraphics[scale=0.6]{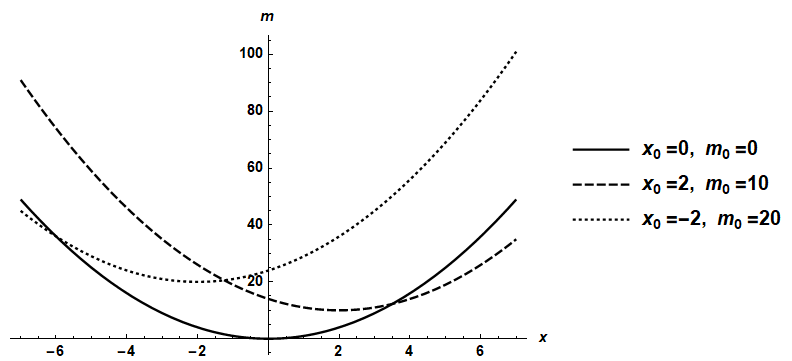}}
     \label{SquaredMasses}
    \end{figure}

Actually, parameters $x_0$ and $m_0$ allow us to have a diverse population with respect to the position around which they move more freely and how much they oppose to an intended change, respectively.
Figure \ref{SquaredMasses} illustrate three of these individuals through their corresponding parabolic mass functions with distinct parameters $x_0$ and $m_0$.
    
Before moving on, there is an extra consideration coming from this example. 
Note that a parabolic position mass function has domain $(-\infty,\infty)$. 
However, other positively concave functions could be used to model more confined individuals.
Also, in the many scenarios of social phenomena, one could encounter systems where a mass function with negative concavity serves as a better model. 
In such a case, one has a repelling effect around a given conviction fixed by the position of the maximum of the mass function.
For reasons already stated, we expect that concave up mass functions are more natural and frequent in describing convictions.
    
\subsection{Forces acting on individuals}
Having described how to build a model for an particle evolving freely, now, let us investigate the behaviour of a particle subject to a net force $F=F(x,t)$.
Again, we use a simple, yet illustrative, initial model: a modified version of Hooke's law.
The functional form proposed for the force is a linear function:
    \begin{equation}
     F = ar + b \,,
    \end{equation}
where $r=|x-x_c|$ is the distance between the position $x$ of the individual and a certain position $x_c$ around which the force is centered.
Also, $a$ and $b$ are constants to describe the particular form and properties of the force.
Note that this type of forces can be obtained from a potential function
    \begin{equation}\label{potential}
     U = -\left(\frac{a}{2}r^2 + br\right) \,,
    \end{equation}
by means of $F=-\partial_r\,U$.

\begin{figure}[h]
     \caption{Simple model of a potential function $U$ given by equation (\ref{potential}) with parameters $a=2$ and $b=-2$.}
     \centerline{\includegraphics[width=\linewidth]{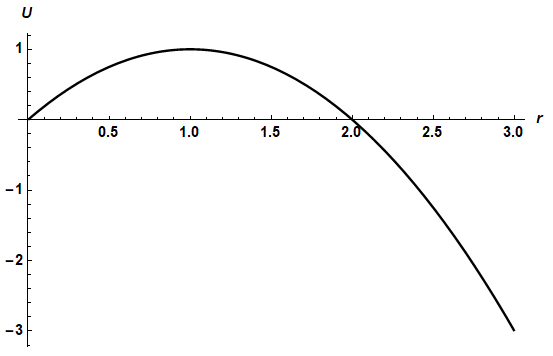}}
     \label{Apot}
    \end{figure}

Figure \ref{Apot} shows an example of a this potential, where $a<0<b$. 
Then, there is an attractive and a repulsive region. 
    
Indeed, the change of sign in $F$ is important. 
The negative part, closest to $r=0$ is the region where \emph{affinity} or proximity of the individual to the position $s_c$ produces an attractive force, thus, a negative force as is conventional in physics.
On the other hand, in this model, there is a single point where the force is null, $F=0$, after which the distance $r$ between the individual's position to the centre $s_c$ yields a repulsive force.
Due to these properties, one can note that, although simple,this model is a useful first approximation to human interactions.
        
Since the force is linear in $r$, $a$ describes the overall strength of the force, whether in the attractive or repulsive regions, while $b$ describes the maximum of the attractive force and, for a fixed value of $a$ the size of the attractive region. 
Certainly this is a very simple model of a force.
More complicated models can be used, but $F=ar+b$ is a convenient starting point to illustrate the possibilities of the ideas proposed here.
    
For the current example, with the simplifying assumption $x_c=0$, the equation of motion becomes 
        \begin{equation}
         x^2\,{\rm d}_t^2x + 2x({\rm d}_tx)^2 = ax + b \,,
        \end{equation}
whose general solution is implicitly given by
        \begin{equation}
         \int \frac{x^2}{\sqrt{3ax^4+4bx^3+6C_0}}\,{\rm d}x = C_1 \pm \frac{t}{\sqrt{6}} \,,
        \end{equation}
where $C_0$ and $C_1$ are constants to be determined from the initial conditions.
    
\begin{figure}[h]
     \caption{Directional field of a motion with quadratic position-dependent mass, under a force $F=2(x-1)$,  compatible with initial conditions $x_i=1/2$ and $v_i=1$.}
     \centerline{\includegraphics[width=\linewidth]{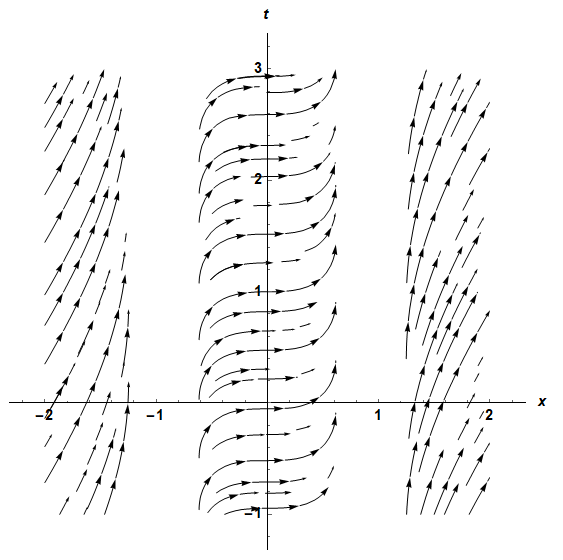}}
     \label{MovFor1}
    \end{figure}

\begin{figure}[h]
     \caption{Directional fields of a motion with quadratic position-dependent mass, under a force $F=2(x-1)$,  compatible with initial conditions $x_i=1/2$ and $v_i=3$.}
     \centerline{\includegraphics[width=\linewidth]{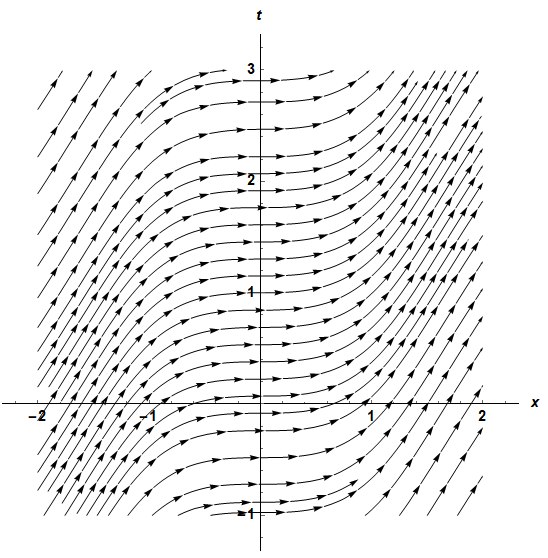}}
     \label{MovFor2}
    \end{figure}

To develop some intuition, consider the directional fields in  Figures \ref{MovFor1} and \ref{MovFor2} describing two types of motion resulting from this model.
The direction forward in time is upward along the vertical axis.
Without loss of generality, motion is assumed from left to right. 
In the case of Figure \ref{MovFor1} the conditions are not enough for the individual to escape the attractive and repulsive regions.
These are the domains where the force $F$, that is, the derivative of $U$, takes negative and positive values, respectively; see Figure \ref{Apot}). 
In the case of Figure \ref{MovFor2} the conditions are \emph{strong} enough so that the individual moves through the different regions.
    
The motion just presented can describe an individual's position, stance, belief, conviction in the presence of the given external force.
However, such a force could be produced by another particle, thus providing a model for an interaction between them.
Note, also, that such an interaction need not be symmetrical, like in Newton's third law.
In this framework for social mechanics we have a rich set of possible models: 
two particles could have reciprocal interactions, but they could respond differently to it due to differences in their mass functions; or they could have no reciprocal interactions at all.

Thus far, quantities have possessed well determined values. 
However, the aggregation of particles usually results in an intractable set of equations.
Now, we proceed to study how stochasticity can be included in our framework.
    
\subsection{Stochastic motion}
Let us proceed now to the case where the influences acting on a social particle include a stochastic term.
Thus, a practical uncertainty is added to the laws governing its motion.
This can help make assertions about social behaviour even when interactions seem intractable. 
This is also in a closer resemblance to the many ways in which individual stances on social issues are influenced by many random signals in the social ambience.
    
First, to account for the added uncertainty, instead of using single position, we consider a distribution of positions $f=f(x,t)$.
In a probabilistic interpretation, once normalized, $f(x,t)$ can be used to obtain the probability of finding an individual in position $x$ at time $t$.
    
While there are many distributions used to describe populations in social studies, the normal distribution is probably the most frequently used to describe them, even if in an approximate fashion.
Thus, let us use it to show how the time evolution of the beliefs of a population could be treated in terms of particle position distributions.
    
Consider the Smoluchowski equation 
    \begin{equation}\label{FPeq}
     \partial_t f = \partial_x^2\left(\alpha f\right) - \partial_x\left(\beta f\right) \,,
    \end{equation}
where, in general, $\alpha=\alpha(x,t)$ and $\beta=\beta(x,t)$ are called the \emph{diffusion} and \emph{drift}, respectively.
This equation can be obtained from the overdamped Brownian motion of a particle, i.e. a particle subject to a force $F(x,t)=\gamma\,{\rm d}_tx+\beta+\sqrt{2\alpha}\,\eta(t)$ with negligible inertial term $m\,{\rm d}_t^2x$.
Here, $\eta$ is a stochastic function usually motivated by random collisions.

Thus, equation (\ref{FPeq}) describes the time evolution of the probability density function of a particle under the influence of random and drag forces, described by $\alpha$ and $\beta$, respectively.
In a social mechanical sense, this model describes a population subject to a compelling trend $\beta$ and some noise $\alpha$ that scatters agglomerations of beliefs.
For a simple case, the result is a non-uniform distribution that moves in time changing both its mean, as well as its variance.
Let us see how.
    
Suppose that both the diffusion and drift are functions of time only: $\alpha=\alpha(t)$ and $\beta=\beta(t)$.
Thus, the external effort to shift the stance of the population and the noise preventing uniformity act equally across all stances, although it varies in time.
Under this simplifying assumptions, the well known solution of (\ref{FPeq}) is a normal distribution
    \begin{equation}\label{soln}
     f(x,t) = A\,{\rm e}^{-[x-\mu]^2/2\sigma^2} \,,
    \end{equation}
with time dependent mean $\mu=\mu(t)$ and variance $\sigma^2=\sigma^2(t)$, and $A$ a normalization factor.
The parameters of the distribution can be obtained from those of (\ref{FPeq}) by
    \begin{equation}\label{timecoeff}
     \partial_t \mu = \beta \,, \quad \partial_t \sigma^2 = 2\alpha \,.
    \end{equation}
That is to say, at each moment in time the stances of the population are described by a normal distribution, the only differences from moment to moment are the parameters $\mu$ and $\sigma^2$ that characterize them.
Even more, the forces act in a transparent way since the drift (overall trend) influences the centre of the distribution while the diffusion (noise) affects the dispersion of the distribution.
    
As a concrete model consider that the diffusion and drift coefficients are oscillations with distinct frequencies $\omega_\alpha$ and $\omega_\beta$, respectively. 
Then, we can write these coefficients in the following form:
    \begin{equation}
     \alpha = \frac{\omega_\alpha}{2}\sin(\omega_\alpha t) \,,
     \qquad 
     \beta = -\omega_\beta\sin(\omega_\beta t) \,.
    \end{equation}
The solution of equation (\ref{FPeq}) corresponding to these coefficients is then given by
    \begin{equation}
     f(x,t) = \frac{{\rm e}^{\frac{[x-\cos(\omega_\beta t)]^2}{2 (\cos[\omega_\alpha t)-2]}}}{\sqrt{2\pi[2-\cos(\omega_\alpha t)]}} \,.
    \end{equation}
Indeed, this is a normal distribution whose mean $\mu=\cos(\omega_\beta t)$ and variance $\sigma^2=2-\cos(\omega_\alpha t)$ oscillate with frequencies $\omega_\alpha$ and $\omega_\beta$, respectively. 
    
Other models can be produced in a similar fashion by means of (\ref{timecoeff}), given the validity of assuming that diffusion and drift depend on time only.
Even more models could be obtained by dropping these assumptions.

To illustrate how the proposed framework can be connected with empirical data, consider the following simplified representation of partisan preferences in the United States. 
The objective of this example is not to validate a unique microscopic
model of political behaviour, but rather to demonstrate how aggregate electoral outcomes may be embedded into an effective dynamical description in terms of collective variables such as drift and diffusion.

Consider the outcomes of the United States of America (USA) presidential elections since 1856. 
We have selected this period since then the two party system is clearer, with democrats and republicans winning either the first and second place, or vice versa, in majority of the popular vote; except for the election of 1912, where the Progressive Party came in second.
    
The configuration of the model is as follows:
We focus only on the population that voted either for the Democrat or the Republican party.
All else are considered as external to this particular system.
The coordinate $x$ describes the position of an individual in a one-dimensional stance-space with respect to both parties.
That is to say, the position space is described by the real line $\mathbb{R}$, where the more to the left the position is ($x<0$), the more the individual supports the Democrat party; the more to the right the position is ($0<x$), the more the individual supports the Republican party; and $0$ is for an indifferent stance with respect to both parties.
    
Next, we suppose that the evolution of the population can be approximated by the model described by equation (\ref{FPeq}), where the diffusion and drift are time dependent only, so that the distribution of preferences in the population is described by equation (\ref{soln}).
Hence, even though all the data we have is the percentage of people that voted democrat or republican, the model implies a normal distribution of preferences within the population compatible with the outcome of each election. 
The resulting means and variances of the elections can be seen in Table \ref{MeanVar}.

    \begin{table}[!ht]
    \centering
    \caption{Mean and variances of normally distributed preferences compatible with the outcomes of USA's presidential elections under the model described by equation (\ref{soln}).}
    \label{MeanVar}
     \begin{tabular}{@{}lrrclrrclrr@{}}
      Year & Mean & Variance & ~ & Year & Mean & Variance & ~ & Year & Mean & Variance  \\ 
      1856 & -0.199 & 1.014 & ~ & 1912 & -0.373 & 1.016 & ~ & 1968 & 0.011 & 1.067  \\ 
      1860 & 0.195 & 1.054 & ~ & 1916 & -0.053 & 1.300 & ~ & 1972 & 0.305 & 1.018  \\ 
      1864 & 0.151 & 1.193 & ~ & 1920 & 0.361 & 1.018 & ~ & 1976 & -0.032 & 1.214  \\ 
      1868 & 0.068 & 1.025 & ~ & 1924 & 0.406 & 1.038 & ~ & 1980 & 0.145 & 1.085  \\ 
      1872 & 0.198 & 1.323 & ~ & 1928 & 0.233 & 1.047 & ~ & 1984 & 0.242 & 1.043  \\ 
      1876 & -0.046 & 1.215 & ~ & 1932 & -0.241 & 1.043 & ~ & 1988 & 0.114 & 1.167  \\ 
      1880 & 0.001 & 1.143 & ~ & 1936 & -0.331 & 1.041 & ~ & 1992 & -0.092 & 1.057  \\ 
      1884 & -0.008 & 1.039 & ~ & 1940 & -0.167 & 1.327 & ~ & 1996 & -0.125 & 1.051  \\ 
      1888 & -0.014 & 1.268 & ~ & 1944 & -0.103 & 1.089 & ~ & 2000 & -0.007 & 1.047  \\ 
      1892 & -0.044 & 1.029 & ~ & 1948 & -0.069 & 1.168 & ~ & 2004 & 0.034 & 1.088  \\ 
      1896 & 0.064 & 1.166 & ~ & 1952 & 0.143 & 1.042 & ~ & 2008 & -0.096 & 1.039  \\ 
      1900 & 0.081 & 1.030 & ~ & 1956 & 0.201 & 1.028 & ~ & 2012 & -0.052 & 1.049  \\ 
      1904 & 0.261 & 1.029 & ~ & 1960 & -0.002 & 1.110 & ~ & 2016 & -0.030 & 1.074  \\ 
      1908 & 0.124 & 1.094 & ~ & 1964 & -0.296 & 1.028 & ~ & 2020 & -0.059 & 1.030  \\ 
     \end{tabular}
    \end{table}
   
To illustrate this, consider the distributions depicted in Appendix \ref{App1}.
Distributions for years 1872, 1912, 1924 and 1940 possess particularly contrasting means, variances and overall outcomes.
In each case, the popular vote outcome for the democrat and republican candidate is well approximated by the portion of the population to the left and right of the position of indifference $x=0$, respectively.
    
\begin{figure}[h]
     \caption{Evolution of drift and diffusion for a simple model of USA's two party preferences.}
     \centerline{\includegraphics[width=\linewidth]{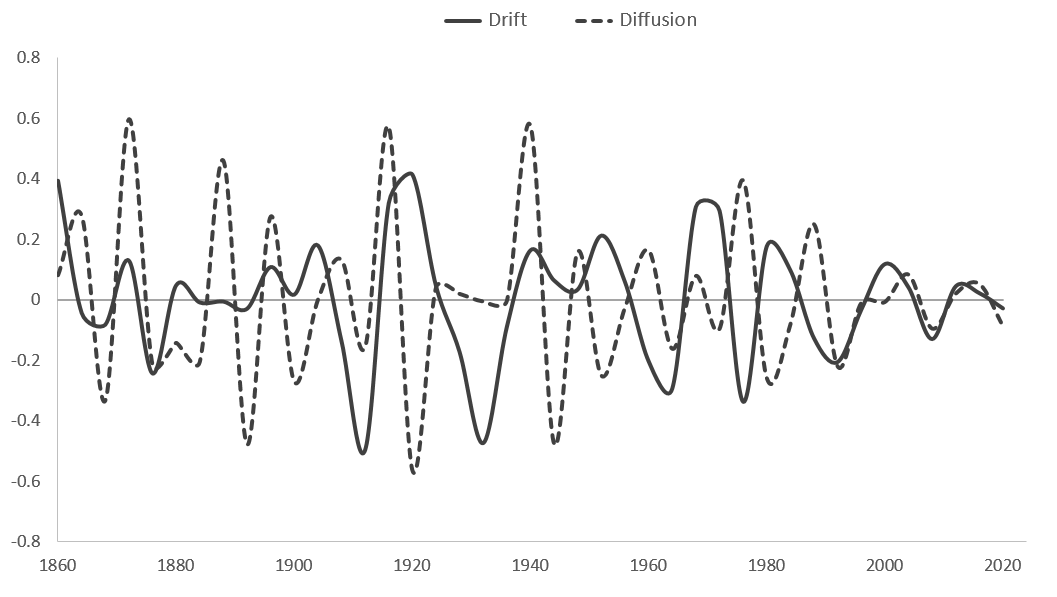}}
     \label{DrifFusion}
    \end{figure}

Finally, the resulting modeling can be used to describe the forces acting on the population, namely, those related to the drift and diffusion of partisan stances.
Assuming they can be approximated by continuous functions, one can use (\ref{timecoeff}) to obtain the drift and diffusion coefficient. 
Numerically approximating them yields the graphs shown in Figure \ref{DrifFusion}.
Therefore, in the framework proposed here, they characterize the effective forces acting on the American society towards modifying its partisan preferences.
We call them effective because one can imagine that these forces are the net result of various social interactions, efforts and compels of diverse nature, whether deliberate or not.
    
It is worth noticing that distribution (\ref{soln}) is unimodal, in line with the hypothesis of the \emph{median voter model} \cite{b48}.
However, in the light of strongly politically polarized societies, bimodal distributions have shown to be effective in describing elections \cite{jsf22,mdb95}.

Admittedly, this is are rather simplified models of the changes in political stances in the USA, but very descriptive initial ones, nonetheless.
These models also show that the framework does not explain the nature of the interactions. 
Although, it may shed some light on certain processes and, more specifically, on effective phenomena.

As mentioned earlier, the Smoluchowski equation (\ref{FPeq}) governs the probability density of finding a particle in the position $x$ at time $t$ given that it is subject to an overdamped Brownian motion. 
However, we wish to include considerations proposed in subsection \ref{parmot}, namely, 
inertial terms of the form $m\,{\rm d}_t^2x + ({\rm d_x}m)({\rm d}_t x)^2$.
Then, consider the motion of an agent whose inertial response depends on its location in stance-space subject to a random force $\eta(x,t)$:
    \begin{equation}\label{sKPZ}
     m\,{\rm d}_t^2x + m'({\rm d}_t x)^2 = \eta \,,
    \end{equation}
where $m'={\rm d}_x m$.
This is the stationary Kardar-Parisi-Zhang equation for $x$ and $t$ playing the role of position $x$, instead.
It has an associated Fokker-Planck equation, similar to the Smoluchowski equation, that admits a solution of the form \cite{rw19}:
    \begin{equation}\label{KPZFPsoln}
     f(t) \sim {\rm e}^{-\frac{m}{2D}\int ({\rm d}_t x)^2 {\rm d}t} \,,
    \end{equation}
where $D$ is a diffusion coefficient that captures the effects of the random noise.
Notably, $f(t)$ does not depend on $m'$.
Expression (\ref{KPZFPsoln}) gives the probability density for a given path $x(t)$.

\section{Other formulations of mechanics}\label{Other}
It is well known that classical mechanics can be formulated in terms of extremal action principles instead of an interplay of forces.
This alternative formulation has helped discover a vast set of features in mechanical systems.
In fact, modern approaches to fundamental interactions in physics are based on these principles.
Let us investigate how similar formulations can contribute to the framework presented here.

The following constructions should be understood as formal extensions of the effective framework introduced previously, rather than as claims about fundamental optimization principles governing human behaviour.

Consider, for example, as an effective variational description, one may suppose that the evolution of an individual's position in the context described here is such that it extremizes a quantity, called \emph{action}:
    \begin{equation}
     S = \int L\,{\rm d}t \,,
    \end{equation}
where
    \begin{equation}
     L = \frac{1}{2}mv^2 - U \,,
    \end{equation}
$m=m(x)$, $U=U(x)$ and $v=\frac{{\rm d}x}{{\rm d}t}$.
Then, $L=L(x,v;t)$ and the evolution of $x=x(t)$ is governed by the functional condition
    \begin{equation}
     \delta S = 0 \,, 
    \end{equation}
subject to a constraint $F_c=-\frac{1}{2}v^2\partial_xm$. 
So that $x(t)$ satisfies the constrained Euler-Lagrange equation
    \begin{equation}
     \frac{{\rm d}}{{\rm d}t}\left(\frac{\partial}{\partial v}L\right) = \frac{\partial}{\partial x}L + F_c \,, 
    \end{equation}
that, in turn, yields eq. (\ref{2ndlaw}) with $F=-\frac{\partial}{\partial x}U$.
The extra term $F_c$ is commonly known as a \emph{thrust} force \cite{r20}.
    
Equation (\ref{2ndlaw}) can also be presented as a constrained Hamiltonian system satisfying Hamilton's equations
    \begin{equation}
     \frac{{\rm d}x}{{\rm d}t} = \frac{\partial H}{\partial p} \,, \qquad
     \frac{{\rm d}p}{{\rm d}t} = -\frac{\partial H}{\partial x} + F_c \,,
    \end{equation}
where the momentum $p=\frac{\partial L}{\partial v}=mv$ and the Hamiltonian function is given by
    \begin{equation}
     H = \frac{p^2}{2m} + U \,.
    \end{equation}
Again, the extra thrust term $F_c$ is present. 
    
One can also rewrite this system in a more convenient form by introducing the following effective quantities, as in \cite{r20}:
    \begin{equation}
     H_{\rm eff} = \frac{m}{m_c}\left(\frac{p^2}{2m} + U_{\rm eff} \right) \,,
    \end{equation}
where
    \begin{equation}
     U_{\rm eff} = U - \frac{1}{m}\int U\,\partial_xm\,{\rm d}x \,,
    \end{equation}
and $m_c$ is a constant.
Then, by defining new mass and momentum variables: $\mu=m^2/m_c$ and $\phi=\mu v$, we can write 
    \begin{equation}
     H_{\rm eff} = \frac{\phi^2}{2\mu} + U_{\rm eff}   
    \end{equation}
and the system satisfies the following equations of motion in the common Hamiltonian form:
    \begin{equation}
     \frac{{\rm d}x}{{\rm d}t} = \frac{\partial H_{\rm eff}}{\partial \phi} \,, \qquad
     \frac{{\rm d}\phi}{{\rm d}t} = -\frac{\partial H_{\rm eff}}{\partial x} \,.
    \end{equation}
    
A possible applications of these alternative formulations of mechanics that immediately comes to mind is the connection with utility theory and others similar where the extremization, whether towards a maximum or minimum, is used to describe human behaviour.
In such a case, one could connect the utility with the action, up to a sign.
    
Another potential application is the inclusion of concepts of symmetries and conserved quantities.
Lagrangian and Hamiltonian formulations of mechanics are rather transparent about the role of these concepts in mechanics.
However, it is not clear now what kind of symmetries and conservation laws would be realized by social phenomena.

We shall not explore further these topics here, leaving it for future work. 

\section{Conclusions}\label{CS}
In this work we have explored the idea behind social physics that analogies between some social and physical systems may occur under certain circumstances.
While an analogy can take place as mathematical formulations coincide, this work studies further comparisons.
Thus, physical concepts are employed to describe social change, taking advantage of the well established heuristics behind physics.
That is to say, the goal is not merely to reuse mathematical tools from physics, but to investigate whether mechanical structures can provide effective descriptions of certain forms of social change.

However, before we finish this exposition, a warning is necessary:
as with other frameworks exploiting analogies, the connections made through social mechanics should be treated as effectively describing social phenomena, as opposed to fundamental explanations.
Any statement on the fundamental nature and mechanisms of an attempted ``true'' description of human behaviour must be rigorously tested against robust data, with great skepticism.
People are not machines. 

For the sake of systematicity, the investigation commenced with the a proposal to connect fundamental concepts in physics with notions regarding social change.
Specifically, we explored the concepts of particle, position, space, motion, inertia, mass, momentum, force, interaction and equilibrium.
They key idea is to represent stances on a social question as positions in a configuration space.
 
Once the initial concepts were established in a social change context, we used the resulting connections to describe a particle moving freely and one moving under the influence of a net force.
An important caveat is the generalization of the concept of mass.
On the one hand, effective inertial parameters may, in principle, assume positive or negative values depending on the phenomenological response being modeled.
The first case represents a resistance to motion under an external influence that increases with the magnitude of the mass.
The second case can be understood as a back reaction to the external influence that produces a motion in the opposite direction of it and it also increases with its magnitude.
 
On the other hand, the mass is also generalized to be a function of position.
With it, we can distinguish the responses to social influences and characterize the social unit depending on its current stance on the topic.
This corresponds to assuming that identical positions in stance-space yield identical effective inertial responses for a given individual.
Although one can imagine more general mass functions, like one that is also explicitly time dependent or with hysteretical behaviour, these cases are not explored here. 

Notably, here we used the Newtonian formulation of classical mechanics.
Thus, a remark on well-known Newton's laws is at hand.
As in the first law, there is a tendency to maintain a particular motion when no forces act on an object: free motion; as well as a measure of this tendency: inertial mass.
However, these differ from the physical case since, although the mass is a characteristic of the object, it varies with its position.
Still, one can use the mathematical expression for the second law, stating that the net force on a particle produces a change in the state of motion through the derivative of momentum with respect to time.
Finally, generalizing the third law, while human interactions can be two-way, one can imagine the reciprocity not being identical. 
In this regard, some asymmetries could be incorporated in the model by means of distinct mass functions among interacting particles.
Nonetheless, it seems that this would need to be validated for a given specific phenomenon.

We also briefly described the Lagrangian and Hamiltonian functions corresponding to the proposed framework.
However, as is well known, the resulting system is non-holonomically constrained due to the position dependent mass.
They could be connected to formulations of social behaviour based on extremization of functions like utility theory.
 
Finally, we showed how stochastic considerations can be incorporated by passing to a mechanical description based on distributions instead of trajectories.
As an example, we used the Smoluchowski equation to describe partisan stances corresponding to the outcomes of USA's presidential elections since 1856.

\subsection{Future work}
From the various considerations presented in this work, there are some immediate prospective directions to continue the investigations presented here.
One can study other possible generalizations of mass functions or different specific cases of forces and interactions.
Notably, we have used one-dimensional systems for simplicity. 
However, multiple social issues in the form an additional dimensions can be considered. 
One can also investigate the many-body problem in social mechanics and even develop an statistical mechanics explanation of how individuals aggregate to produce social configurations, analogously to how small particles aggregate to produce solids, liquids or gases, or even more complex states.
    
There are two more possible directions for the kind of research conducted here: social systems analogous to quantum physics and general relativity.
Although seemingly strange, the first is a reasonable direction since quantum mechanics incorporates fundamentally the notion of stochasticity of a single particle. 
The uncertainty is fundamental and we do not know why it must be accounted for in the quantum theory, so it is postulated.
An attempt could be made to see if the results from quantum mechanics yield mathematical tools to deal with uncertainties in an individual's motion, even if they are not fundamental, but only effective due to the unpredictability of human behaviour \cite{h23,aa15}.
    
General relativity offers possibilities in applying analogous reasoning to social systems.
The novelty of general relativity is an intertwine between the nature of the space of positions and the objects moving on it. 
This can help generalize the notion that the positions accessible, i.e. the space, are previously fixed to be occupied by individuals, moving towards a picture in which the space of stances is dynamical and changes as a society evolves.
While the mathematical framework for general relativity is differential geometry, where space-time is a manifold, there are already social stances describable in these terms, for example, the concept of \emph{equilibrium manifold} in economics \cite{b09}.
Of course, no connection between them is claimed now, but they are susceptible of investigation for possible analogies.
It is worth mentioning here that there is already an empirical result in economics known as the \emph{gravity model of international trade} stating that ``the volume of trade between two countries is proportional to their economic mass and a measure of their relative trade frictions'' \cite{bs20}. 
Although, it is analogous to non-relativistic gravity.
    
In any of these prospectives there is always the opportunity to find data susceptible to be described by each resulting model that, in turn, can help to test and tune it.

\pagebreak

\appendix
\section{Selected examples of the distribution of partisan preferences in presidential elections in the US}\label{App1}
 
 \begin{figure}[h]\label{1872}
    \caption{1872}
    \centerline{\includegraphics[width=\linewidth]{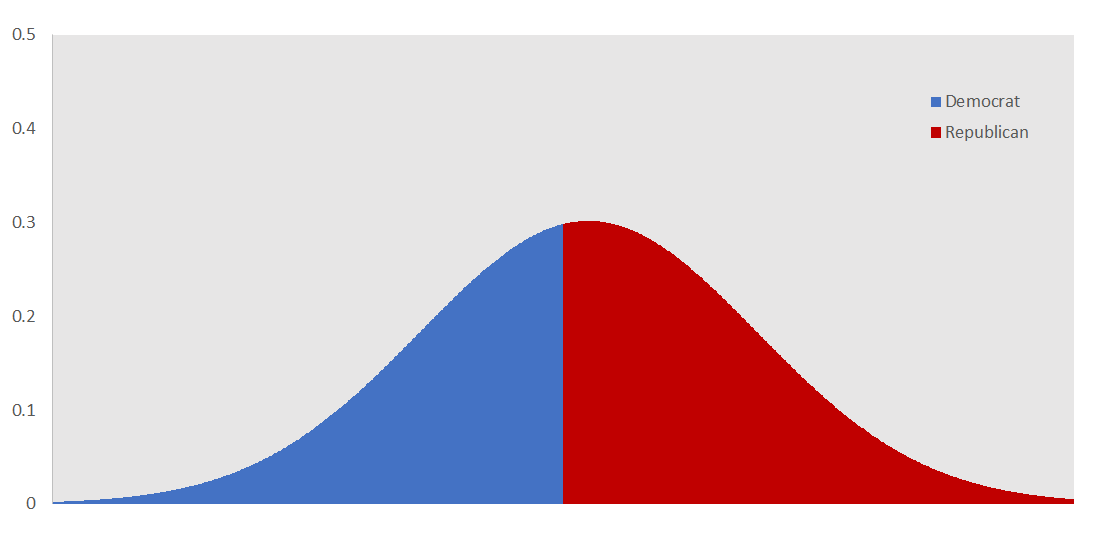}}
  \end{figure}
  \begin{figure}[h]\label{1912}
    \caption{1912}
    \centerline{\includegraphics[width=\linewidth]{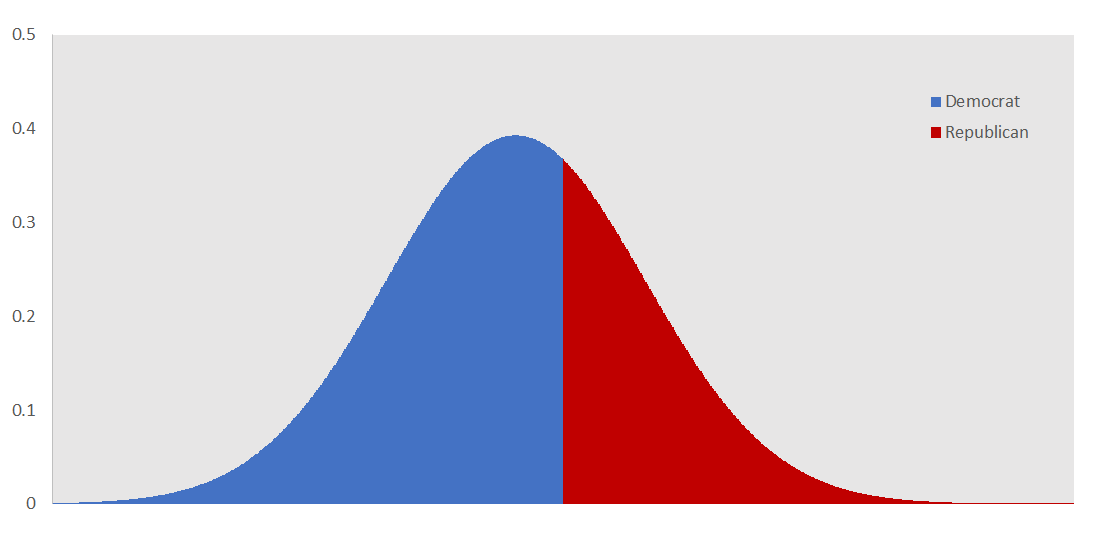}}
  \end{figure}
  \begin{figure}[h]\label{1924}
    \caption{1924}
    \centerline{\includegraphics[width=\linewidth]{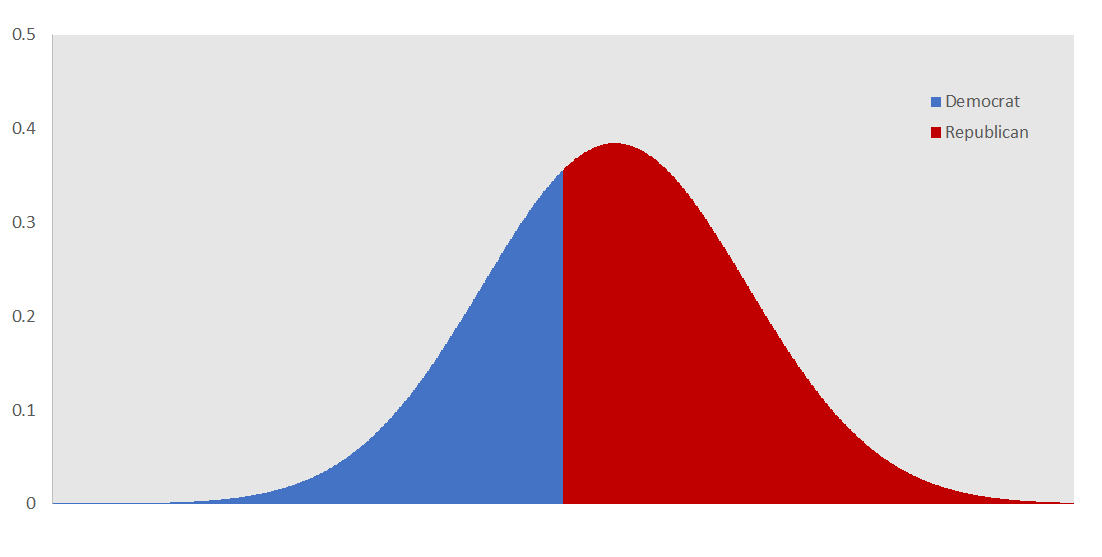}}
  \end{figure}
  \begin{figure}[h]\label{1940}
    \caption{1940}
    \centerline{\includegraphics[width=\linewidth]{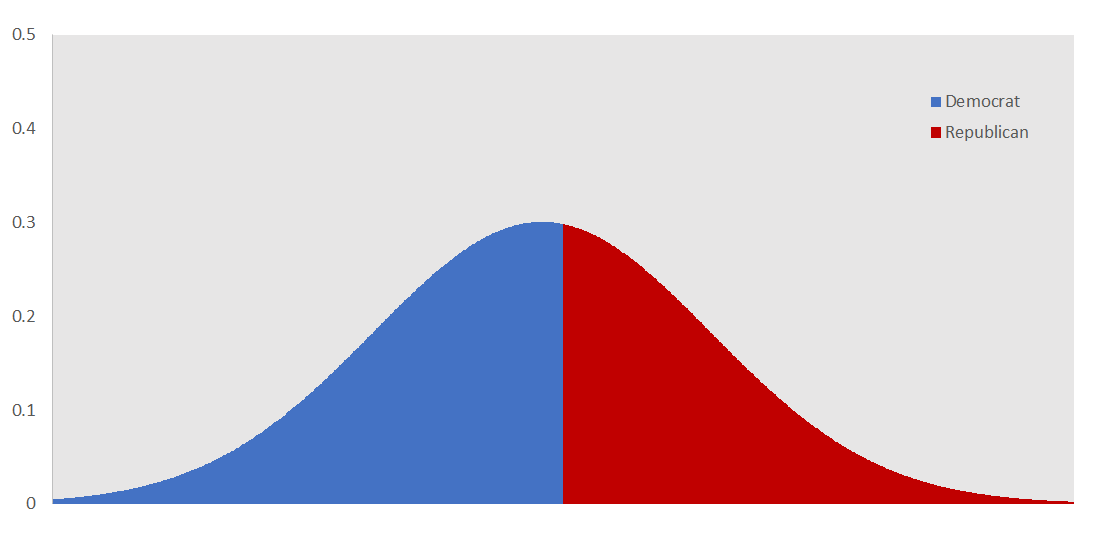}}
  \end{figure}

\end{document}